# Evolution of parsec-scale jet directions in active galaxies

I. M. Kostrichkin ⓘ,[1]★ A. V. Plavin ⓘ,[2] A. B. Pushkarev ⓘ[3,4] and M. S. Butuzova ⓘ[3]

[1]*Moscow Institute of Physics and Technology, Institutsky per. 9, Moscow Region, Dolgoprudny 141700, Russia*
[2]*Black Hole Initiative at Harvard University, 20 Garden Street, Cambridge, MA 02138, USA*
[3]*Crimean Astrophysical Observatory, 298409 Nauchny, Crimea, Russia*
[4]*Astro Space Centre of Lebedev Physical Institute, Profsoyuznaya 84/32, Moscow 117997, Russia*

**ABSTRACT**
We analyse the variability of the parsec-scale jet directions in active galactic nuclei (AGNs). Our analysis involves 317 AGNs at frequencies ranging from 2 to 43 GHz, and is made possible by developing an automatic jet direction measurement procedure. We find strong significant variations in a one quarter of these AGNs; the effect is likely ubiquitous, and not detected in the rest due to a limited sensitivity and observations epoch coverage. Apparent jet rotation speeds range from 0.21 deg yr$^{-1}$ at 2 GHz to 1.04 deg yr$^{-1}$ at 43 GHz. This strong frequency dependence indicates that the variability cannot be explained by jet components propagating ballistically without acceleration: more complex jet shapes or motion patterns are required. Still, we demonstrate that the apparent direction changes are predominantly caused by the jet nozzle rotations, and not by individual components propagating transversely to the jet. In this work, we focus on variability scales much longer than the times of observations, that is $\gtrsim 50$ yr. Using our measurements, we bound potential periods to less than 1000 yr in the source rest frame for 90 per cent AGNs in the sample. These time-scales constrain jet direction variation mechanisms, with the most likely explanations being the plasma instabilities, the precession caused by the accretion disc with density $\sim r^{-1}$, and the orbital motion of binary systems.

## 1 INTRODUCTION

Jets of active galactic nuclei (AGNs) are some of the most energetic non-transient objects in the Universe and generate radiation across almost the entire range of the electromagnetic spectrum. A relativistic jet is a flow of mainly electron–positron plasma bursting out of the centre of AGNs with a velocity close to the speed of light (Beresnyak, Istomin & Pariev 1997). The radio emission, primarily generated through the synchrotron process, can extend from subparsec to megaparsec scales (Blandford & Königl 1979). The submillisecond resolution attained by radio interferometers enables jet geometry studies at scales of parsecs, where the flow is highly relativistic and collimated.

Among AGNs with bright parsec-scale emission, blazars form the majority: the typical jet viewing angles are only a few degrees from the line of sight (Lister et al. 2019). At these viewing angles, any small direction changes of the jet are amplified, and can manifest themselves as strong apparent jet bends or the temporal evolution of the apparent jet direction. The latter has been the focus of many observational studies, both for individual blazars (Agudo et al. 2012; Butuzova 2018; Cui et al. 2023) and for samples of hundreds of AGNs (Lister et al. 2013, 2021). The significant temporal evolution of the inner jet direction was detected in all those studies. There are multiple scenarios that could explain the observed variability, including disc-driven precession (Sarazin, Begelman & Hatchett 1980), gravitational influence in binary black hole system (Valtonen & Wiik 2012) or instabilities in the jet itself, which can lead to changes in the parsec-scale direction (Nikonov et al. 2023).

Massive systematic observational studies are crucial to further constrain these mechanisms and understand the general picture of the AGN jets. The Monitoring Of Jets in AGNs with VLBA Experiments team (MOJAVE) studied the jet direction variability for 447 AGNs at 15 GHz (Lister et al. 2021). Recently, jet direction measurements were performed for thousands of sources at multiple frequencies (Plavin, Kovalev & Pushkarev 2022), but only the average directions were considered and reported there. See Section 3 for an overview and comparison of those measurements approaches. In this paper, we present the results of our jet direction variability measurements on the basis of VLBI (Very Long Baseline Interferometry) images at four frequencies from 2 to 43 GHz. We attempt to quantify this variability and constrain the potential reasons causing such changes in the jets.

The structure of this paper is as follows. Section 2 provides a description of the source sample and observations utilized in this study. Section 3 details the method we use to determine the direction of the jet on a VLBI image. In Section 4, we pro-vide the results of our analysis and compare with earlier studies such as Lister et al. (2021). Finally, Section 5 summarizes our findings.

## 2 OBSERVATIONAL DATA

Our analysis utilizes VLBI observations carried out at frequencies, ranging between 2 and 43 GHz, sourced from the Astrogeo data base[1] – a collection until the end of 2023. This data base includes both restored images and visibility function measurements, which are the initial interferometric observables. The data base is composed of geodetic VLBI observations (Petrov et al. 2009; Piner et al. 2012; Pushkarev & Kovalev 2012), the VLBA[2] calibrator survey (Beasley et al. 2002; Fomalont et al. 2003; Petrov et al. 2005, 2006, 2008; Kovalev et al. 2007), and other VLBI observations, including the results of Helmboldt et al. (2007), Petrov (2011, 2012, 2013, 2021), Petrov et al. (2011a, b, 2019), Schinzel et al. (2015), Jorstad et al. (2017), Shu et al. (2017), Lister et al. (2018), and Popkov et al. (2021).

Since our main interest is to study the variability in the direction of the parsec-scale AGN jets, we have selected for further study only those sources that have more than 10 epochs covering more than 10 yr at a given frequency. In total, 440 AGNs satisfy these conditions (see Fig. 1 for a breakdown by frequency).

The target sources were radio-luminous active galaxies with VLBI flux densities varying from few millijanskys to tens of janskys. The sample is predominantly made up of blazars, that is AGNs with a viewing angle of a few degrees. For an in-depth study of a complete flux density-limited sample within the MOJAVE program at 15 GHz, refer to Lister et al. (2019).

## 3 JET DIRECTIONS

Different ways of determining the direction of the parsec-scale jet for large samples of AGNs have already been utilized in the literature. The largest study in terms of the number of sources is Plavin et al. (2022), where the jet direction estimation is based on a model-fitting approach. The jet position angle (PA) is calculated as the direction from one component to the other, using the two-Gaussian model fitted to visibilities. There, the authors have found that the approach results in robust average jet direction measurements but does not reliably trace the variability. In Kovalev, Petrov & Plavin (2017), Pushkarev et al. (2017), and Plavin, Kovalev & Petrov (2019), the jet direction is determined from the direction of the inner jet 'ridge line', directly from the VLBI images. Defining a reliable ridge line requires a smoothing step, typically performed by fitting a spline. This can pose challenges to measuring the jet direction at a consistent distance over time, especially when individual bright components have emission gaps between them.

In the MOJAVE team studies (Lister et al. 2021), the direction of the jet was determined by the position of the innermost component of the Gaussian model with respect to the core. Another approach was used in Blinov et al. (2020): there, the jet direction was determined as the peak of the intensity angular distribution in the image, with manual processing afterwards. These two methods are fundamentally different, the former uses visibilities, the latter uses images. However, both fundamentally include manual steps. Our goal is to develop a fully automated approach, easily and uniformly applicable to thousands of observations at different frequencies.

---

[1] http://astrogeo.smce.nasa.gov/vlbi_images/
[2] Very Long Baseline Array of the National Radio Astronomy Observatory, Socorro, NM, USA.

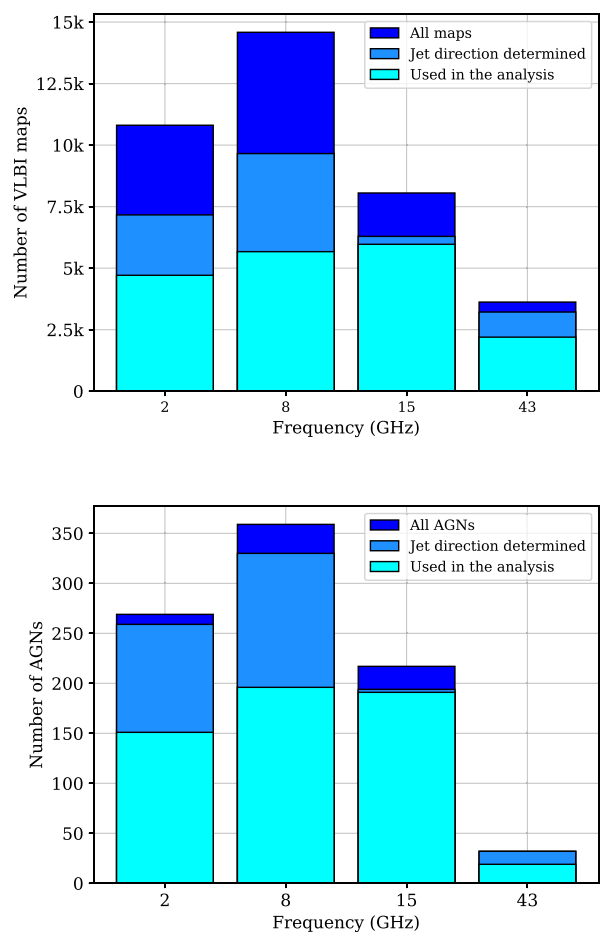

**Figure 1.** Distribution of VLBI images (top) and individual AGNs (bottom) by observing frequency. Dark blue – all available in the Astrogeo data base for AGNs with at least 10 yr of coverage (Section 2); blue – images with jet directions measured in our analysis (Section 3.1); aqua – AGNs with jet direction variability reliably described in our analysis (Section 3.2).

### 3.1 Direction measurements

In this work, we propose a fully automatic method for determining the direction of the inner parsec-scale jets from VLBI images. There are tens of thousands of VLBI images available at a wide range of observing frequencies and epochs (Fig. 1), and any manual approaches are hardly feasible at this scale. We aim to measure the jet directions as close as possible to the origin to track the strongest variations. The distance from the origin should also remain consistent over time for cleaner variability studies.

First, we convolve all images with a circular beam instead of an elliptical one native to the CLEAN algorithm. The main purpose of this substitution is to get rid of the potential difficulties associated with determining the direction of the jet on images with a substantially elliptical beam (Pushkarev et al. 2017). The circular beam size $r$ is chosen to be between the minor $m$ and major $M$ axes of the native ellipse, closer to the minor one: $r = 3/4 \cdot m + 1/4 \cdot M$. Empirically, we see that sizes slightly smaller than the average $(m + M)/2$ preserve image reliability in the high-SNR regions close to the core.

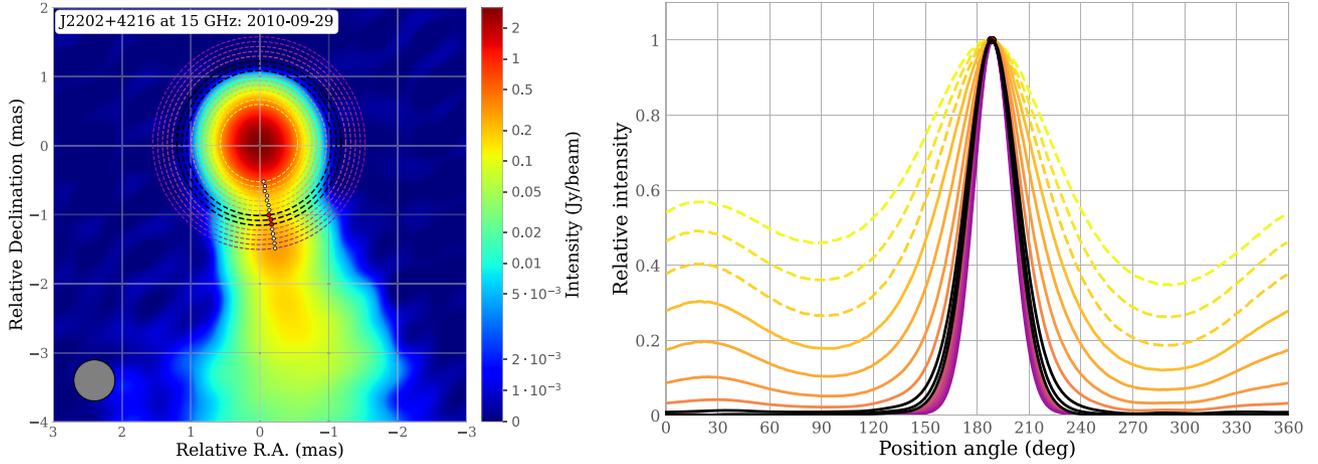

**Figure 2.** Left: Image of BL Lac, obtained on 2010 September 29, at 15 GHz. Dashed circles correspond to distance of 1.6 to 4.4 of the CLEAN beam (FWHM shown in the left bottom corner) from the apparent jet origin. Dots indicate the direction of the jet at these distances. Right: The angular intensity profile ($I/I_{max}$) at corresponding distances from the apparent jet origin in the VLBI image. The jet PA measurement is determined from the peaks of the black curves (Section 3). On the first three dashed curves, the direction of the jet cannot be reliably determined. Thus, the jet direction is determined on the further circles. The PA is measured north through east.

**Table 1.** Summary of observational data properties and jet rotation speed measurements for each frequency band. Columns are as follows: (1) rounded frequency of observation; (2) number of sources used in the analysis on this frequency (see Fig. 1); (3) number of AGNs with significantly variable jet PA on this frequency ($a \geq 3\sigma_a$); (4) distance at which the jet direction is determined (see Section 3); (5) median jet rotation speed (see Section 4).

| Frequency (GHz) (1) | $N_{AGN}$ (2) | $N_{variable}$ (3) | Distance (mas) (4) | Median jet rotation speed (deg yr$^{-1}$) (5) |
|---|---|---|---|---|
| 2 | 143 | 15 | $6.60 \pm 0.35$ | $0.21 \pm 0.02$ |
| 8 | 193 | 44 | $1.85 \pm 0.1$ | $0.27 \pm 0.02$ |
| 15 | 180 | 79 | $1.05 \pm 0.07$ | $0.33 \pm 0.03$ |
| 43 | 19 | 12 | $0.40 \pm 0.02$ | $1.06 \pm 0.15$ |

Second we identify the pixel on the image that has the highest intensity, hereafter we will associate this pixel with the core (apparent origin of the jet) on the image. Afterwards, we construct circles centred at the core with the size ranging from 1.6 to 4.4 times the full width at half-maximum (FWHM) size of the beam. This range was systematically covered in 15 steps: each successive circle increases by $0.2 \cdot FWHM_{beam}$ as illustrated in Fig. 2.

Third, on each circle, we determine the pixel having the highest intensity and, calculate the difference in coordinates between this pixel and the apparent start of the jet. Then, using the obtained vector we calculate the PA of the jet as demonstrated in Fig. 2. Each individual curve corresponds to an angle distribution of intensity at a certain distance. The colour indicates the distance from the apparent jet origin: brighter – closer to the apparent origin of the jet. After this procedure, for each individual VLBI image, we obtain a set of defined PAs at different distances.

We consider the jet direction to be reliably determined at a certain distance from the core when the ratio of the maximum to median intensity on the corresponding curve is greater than $I_{max}/|I|_{median} \geq 3.25$. This cut-off level is chosen empirically and separates two limiting cases: (i) the intensity is almost-uniformly distributed at the target distance with no extended jet visible, and (ii) the emission peak is highly prominent in the direction of the jet. For consistent variability evaluation, we measure the jet direction at a fixed core separation for each observing frequency. Specifically, we determine the closest separation where the jet direction is reliably determined for 90 per cent of images at a given frequency and use it for all images (see Table 1); this is visualized by the middle of the three black circles in Fig. 2. Further, we estimate the direction uncertainty $\sigma_{PA}$ by taking measurements at core separations within $\pm 0.2 \cdot FWHM_{beam}$ from the base value, and taking half of the minimum-maximum range of these measurements. This separation interval corresponds to the three highlighted circles in Fig. 2. Below, we only utilize epochs with $\sigma_{PA} < 45°$; see Fig. 1 for statistics on the number of images and sources.

Compared to earlier jet direction variability studies, our approach is fully automatic. This accelerates the processing speed and makes systematic studies of thousands of images feasible. Also, we design the procedure to measure the jet direction at a consistent distance from the core, that is important to consistently trace the evolution. As a result, after processing the individual images and determining the projected jet direction, 417 AGNs remain from the initial sample of 440 (Fig. 1). Note that the component-based approach presented in Lister et al. (2021) may perform better if the jets structure on the VLBI image is strongly warped close to the origin; the fraction of strongly curved jets is quite low though, $<10$ per cent (Makeev et al. in preparation).

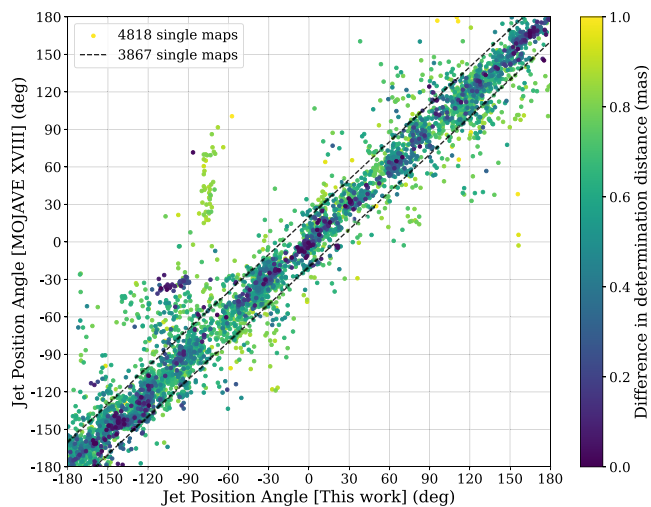

**Figure 3.** Comparison between the PAs measured by MOJAVE (*y*-axis) and those determined by our method (*x*-axis). The colour scale indicates the difference between the distances at which we and MOJAVE determined the direction of the jet on a particular image. The darker the points, the smaller the difference between these distances. A total of 4818 VLBA images were considered, and 3867 of them are in the tube with a spread of less than 20°.

We also performed a comparison of our method for determining the direction of the jets with the method used in Lister et al. (2021). The results of this comparison are shown in the Fig. 3, where the colour indicates the absolute difference between the distances at which the direction of the jet was determined by us and MOJAVE. For almost 80 per cent of the VLBI images at 15 GHz, the difference in the measured jet PA is less than 20°. A separate set of points located in the vertical band from 20° to 80° at −80° horizontal corresponds to quasar J0555 + 3948 (DA 193). According to the MOJAVE results, which determine the direction of the jet as close to the core as possible, the jet direction of this source experiences a significant variations on small spatial scales, while at the scales we analysed the jet direction there are no changes, which leads to the difference in the determined PA between us and MOJAVE.

### 3.2 Variability analysis

At the second stage of the analysis, we work with the obtained PAs. Combining all VLBI observations at a certain frequency of the selected AGNs, we plot the dependence of the jet PA on time. Several examples of the time dependence of the PA is shown in Fig. 4.

In the plots, the epochs of observations are located on the horizontal line, and the values of the jet PA on the vertical line. The purpose of the analysis is to reveal global trends. Therefore, in this work we study changes in the jet direction occurring on scales comparable or significantly larger than the observation time, which for individual AGN is up to 30 yr. For this reason, the behaviour of the jet PA is described by a linear model:

$$PA(t) = a \cdot (t - \langle t \rangle) + PA_0.$$

where $a$ is the jet rotation speed in $\deg\,\mathrm{yr}^{-1}$, $PA_0$ is the jet PA at the mean epoch, and $\langle t \rangle$ is average time over all observation epochs. Unlike more detailed models that involve specific geometrical assumptions (e.g. Butuzova & Pushkarev 2020; Cui et al. 2023), this model describes the behaviour of the PA of most AGNs quite well, and has a minimum of free variables. We assume that the linear dependence reliably captures the behaviour of the jet PA on time if the slope error $\sigma_a \leq 0.5\,\deg\,\mathrm{yr}^{-1}$. Among the considered 417 AGNs, only 317 sources satisfy this $\sigma_a$ criterion at least at one frequency band, see Fig. 1 and Table 1 (the $N_{AGN}$ column).

## 4 RESULTS

### 4.1 Jet direction variability: detections and rates

General information of geometrical jet properties of all sources of the sample is summarized in Table 2. Examples of jet direction measurements and their constant-velocity fits are shown in Fig. 4. We find that $\approx 27$ per cent of the AGNs show significant jet direction variability ($|a| \geq 3\sigma_a$). For the remaining 70 per cent, the apparent variations are within our uncertainties. Still, we use all AGNs in calculating average jet rotation speeds, so that to avoid a bias towards large rotation speeds.

Not for all AGN jets the evolution of their position is well described by a linear trend. We estimate that about 10 per cent of our sample demonstrate sharp jumps and rapid changes in the apparent jet direction; see Section A for more examples and their discussion. Still, the majority of the jets do admit a constant-velocity description on the time-scale of our observations (up to 30 yr).

The full distribution of apparent jet rotation speed for each observing frequency is shown in Fig. 5. The typical speeds are very different for different frequencies: see their medians marked in Fig. 5 and shown together with their errors in Fig. 6. Median speeds range from 0.21 deg yr$^{-1}$ at 2 GHz, to 1.04 deg yr$^{-1}$ at 43 GHz (see Table 1). We observe a tendency for the median value of the jet rotation speed to increase with increasing frequency of observations. Higher frequencies probe regions close to the jet origin efficiently. We interpret this trend as decreasing direction variability further downstream the jet.

In an otherwise uniform sample, one would expect to see differences in the jet rotation speeds between different classes of AGNs observed at different angles, and between AGNs located at different distances. In practice, however, such a comparison turns out to be challenging due to fundamental observational biases: basically, more distant AGNs have to be more luminous and highly beamed. Earlier, Lister et al. (2021) have shown a slightly larger jet PA variations for quasars than for BL Lacs at 15 GHz. We compare jet rotation speeds aggregated in Table 1, and find a similar 15 GHz trend that is barely significant. At other frequencies (2, 8, 43 GHz) though, we do not find any consistent differences of this kind. Currently, we avoid interpreting the interclass differences or lack thereof, but it could be feasible with a cleaner carefully selected subsample of AGNs.

Slower and less pronounced variability downstream implies a jet structure more complex than acceleration-free motion along the cone surface. Indeed, assuming a wobbling jet nozzle and ballistic propagation of components afterwards would have led to a constant jet rotation speed – along the jet and at all observational frequencies. Our results indicate that either the jet shape is non-conical at these scales, or complex component motion patterns also play a role in apparent direction variations. The non-conical jet profiles appear to be quite common in nearby AGNs Asada & Nakamura (2012) and Kovalev et al. (2020), although they appear to transition to a cone closer to the origin than the scales relevant for our work. More complex motion patterns are possible even for the jets that have the conical shape on average. These patterns

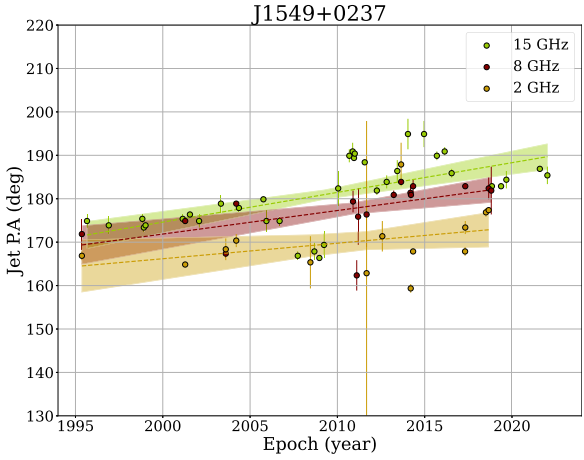

(a) In J1549+0237, the jet appears to be uniformly rotating over time, with consistent trends observed at 2-15 GHz.

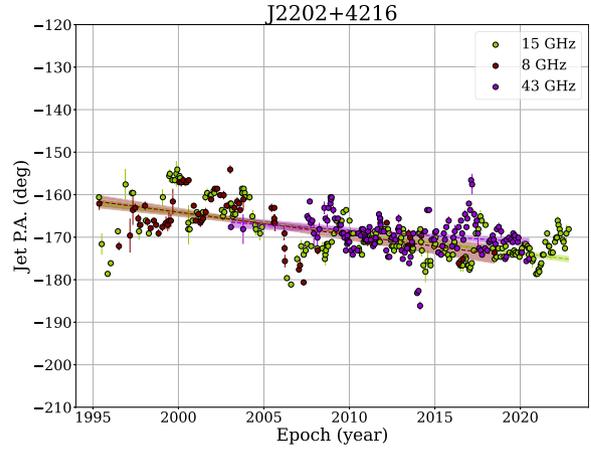

(b) J2202+4216 (BL Lac) is another source with a linear trend with small wobbling on top of that. The direction of the jet varies almost equally for all frequencies. More detailed study of the inner jet is done by Arshakian et al. (submitted).

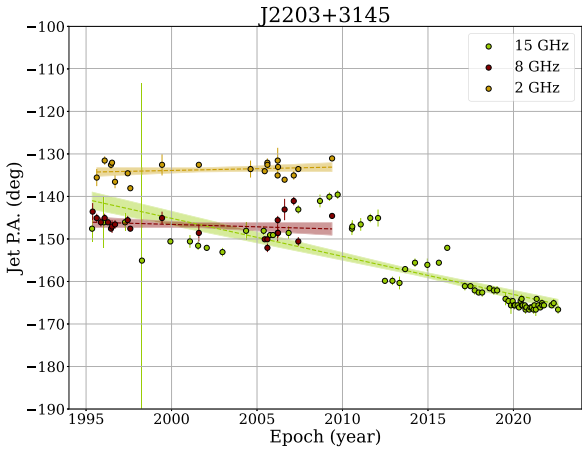

(c) In J2203+3145, jet P.A. variability becomes noticeable only with increasing frequency; at 2 and 8 GHz the direction of the jet is almost independent of time.

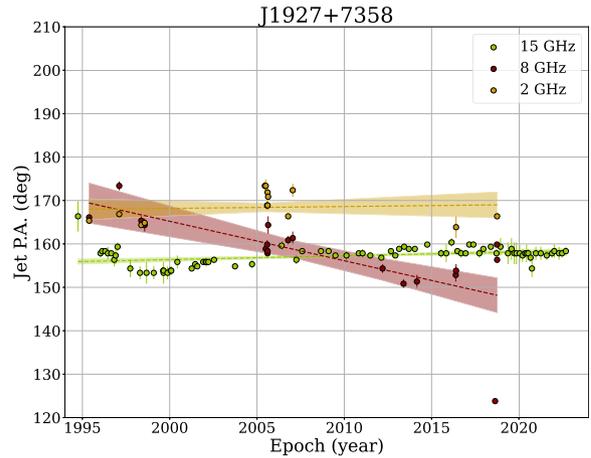

(d) The 15 GHz J1927+7358 jet is an example of a time-independent jet direction ($|a| < 3\sigma_a$)., while apparent variations are seen at a lower frequency of 2 GHz.

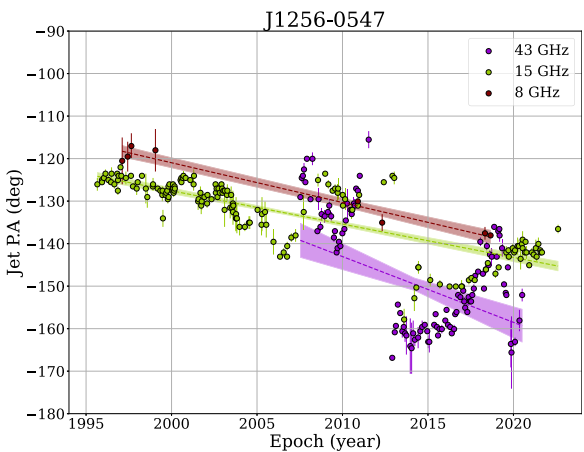

(e) J1256-0547 (3C 279) has a general trend of a steadily rotating jet at 15 GHz, but there is a fairly strong change from 2012 to 2016. The jumps at 43 GHz in 2011 and 2012 are further illustrated in Figure A1.

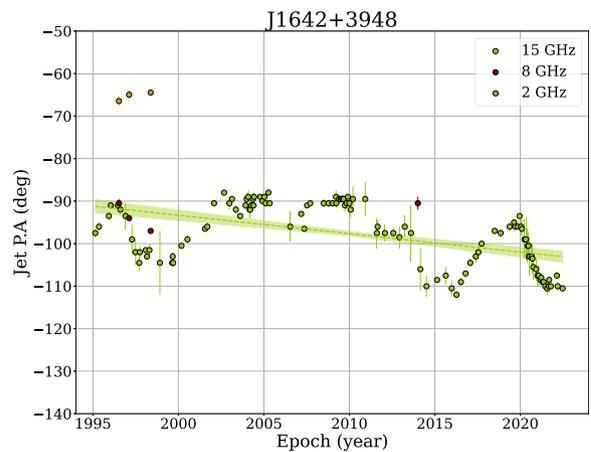

(f) The jet of J1642+3948 (3C 345) demonstrates strong variations on top of the trend; they may even appear quasi-periodic.

**Figure 4.** Plots of the jet PA versus time for six selected AGNs. They show a diversity of global jet direction variability patterns: from the direction staying almost constant in time, to constant rotation speed, to more complex behaviours. The dashed lines shows the result of linear fitting (constant rotation speed assumption, Section 3.2) together with its $1\sigma$ uncertainties. Plots for other sources are available online.

**Table 2.** Measured jet rotation speeds, for each AGN at each observing frequency. Columns are as follows: (1) J2000 Name; (2) number of VLBI epochs contributing; (3) frequency of observation; (4) observation epoch coverage; (5) jet rotation speed in linear fit; (6) mean value of the PA of the jet at the full observation time.The table is published in its entirety in the machine-readable format. A portion is shown here for guidance regarding its form and content.

| J2000 name | $N_{epochs}$ | Frequency (GHz) | Duration (yr) | $a \pm \sigma_a$ (deg yr$^{-1}$) | $PA_0 \pm \sigma_{PA_0}$ (°) |
|---|---|---|---|---|---|
| (1) | (2) | (3) | (4) | (5) | (6) |
| J0006 − 0623 | 57 | 2 | 24.3 | −0.16 ± 0.05 | −78 ± 1 |
| J0006 − 0623 | 53 | 8 | 21.7 | −1.78 ± 0.33 | −75 ± 2 |
| J0013 + 4051 | 22 | 8 | 25.4 | −0.01 ± 0.10 | −30 ± 1 |
| J0017 + 8135 | 81 | 2 | 24.0 | −0.03 ± 0.03 | −168 ± 1 |
| J0017 + 8135 | 10 | 15 | 23.2 | 0.21 ± 0.03 | −2 ± 1 |
| J0019 + 7327 | 19 | 2 | 24.6 | −0.36 ± 0.41 | 170 ± 4 |
| J0019 + 7327 | 11 | 15 | 26.3 | −0.13 ± 0.17 | 129 ± 2 |
| J0038 + 4137 | 11 | 8 | 22.1 | −0.40 ± 0.32 | 97 ± 3 |
| J0057 + 3021 | 15 | 15 | 24.4 | 0.15 ± 0.06 | −51 ± 1 |
| J0102 + 5824 | 74 | 2 | 23.0 | 0.15 ± 0.23 | −106 ± 2 |
| J0102 + 5824 | 109 | 8 | 24.4 | 0.73 ± 0.15 | −123 ± 2 |

include accelerating trajectories of bright jet features, either along the jet or perpendicular to it (Lister et al. 2019), and non-radial motion potentially caused by plasma waves (Cohen & Meier 2015; Cohen et al. 2015) or instabilities (Hardee 2003; Nikonov et al. 2023). Determining and quantifying the exact structure and velocity profile of the jet remains an open question and a subject of further studies.

### 4.2 Direction variability and non-radial motion

The reasons for the apparent rotation of the jet can be different. In this section, we consider the scenario according to which the rotation of the apparent direction of the jet occurs due to the non-radial motion of its individual components. Detailed consideration of the kinematics of individual components has been extensively discussed in, e.g. Lister et al. (2021). Here, we use data from that paper. Each component is described by the mean angular distance from the core feature – $\langle R \rangle$, the speed of proper motion – $\mu$, and its direction, which is given by the offset between mean PA of component and its velocity vector PA, $\langle \nu \rangle - \phi$. We determine the angular velocity of the component relative to the core as

$$\omega = \frac{\mu \sin[\langle \nu \rangle - \phi]}{\langle R \rangle}.$$

In our further analysis, we use one VLBI component for each object: the one with the average distance from the core component closest to 1.05 mas (see Table 1) to determine the angular rotation velocity of the jet. This was done in order to compare the rotation speeds at approximately the same distance from the apparent origin of the jet. There are 419 sources in the MOJAVE sample for which at least one component other than the core is detected, and 191 AGNs participated in our jet direction studies at 15 GHz. In this analysis, we considered only those AGNs for whose components the angular velocity $\omega$ was determined with an error $\sigma_\omega \leq 0.5$ deg yr$^{-1}$.

After applying these selection criteria, 79 AGNs remain. The selected components in these 79 sources exhibit median angular velocity of $\omega_{median} = 0.45 \pm 0.05$ deg yr$^{-1}$. This average is comparable to average jet rotation speeds shown in Fig. 5. The direct comparison of jet rotation and component motion angular velocities for each AGN is presented in Fig. 7. There is no apparent correlation in this plot, we do not see a strong connection between the changes in the apparent direction of the jet and the non-radial motion of its components. The lack of such a connection indicates that the apparent jet direction variability is caused by the effective jet nozzle rotations over time, and not by the transverse motion of its individual components. This way, new bright components in the jet appear in different directions from the start.

### 4.3 Apparent opening angles and direction variability

In order to study the apparent changes in the jet geometry, we also compared the jet apparent opening angles and the amplitude of its PA variations. These quantities are measured with different underlying assumptions, and such a comparison can potentially demonstrate the boundaries of those assumptions. We estimate the PA variations amplitude as the width of the central interval containing 68 per cent individual jet direction measurements for a given source at a given frequency band. The apparent opening angles were measured from stacked VLBI images within the MOJAVE programme (Pushkarev et al. 2017). We ensure that these measurements use the same observational data as we do in our analysis by limiting ourselves to epochs before 2015. For consistency, we take opening angles measured at 1.05 mas from the core: our 15 GHz jet direction measurements take place at this separation (Table 1).

Due to stacking used to measure opening angles, we expect the apparent opening angles to be similar or exceed the amplitude of PA variations. Such a comparison can serve as a consistency check, and can also reveal effects influencing these two measurements in different ways. These results are shown in Fig. 8.

The most notable result of this comparison is a sample of AGNs with $\Delta PA \gg \alpha_{app}$. Such a ratio is possible only if the jet changes its direction drastically, and these changes are not captured by the opening angle stacking. The leftmost top point corresponds to quasar J1224 + 2122 (4C 21.35): its PA changes by more than 70° over 10 yr, while its apparent opening angle is less than 20°. Selected VLBI images demonstrating its rotation are shown in Fig. A1(b). The other two sources that have a $\Delta PA$ almost twice as high as $\alpha_{app}$: J0152 + 2207 and J0217 + 7349. The first has groups of observations in the 1990s and 2020s, between which the direction of the jet changes by about 50°. The second, shows a jump in PA of almost 40° in 2021. Still, the majority of sources have $\Delta PA < \alpha_{app}$, as expected for smooth consistent direction variations.

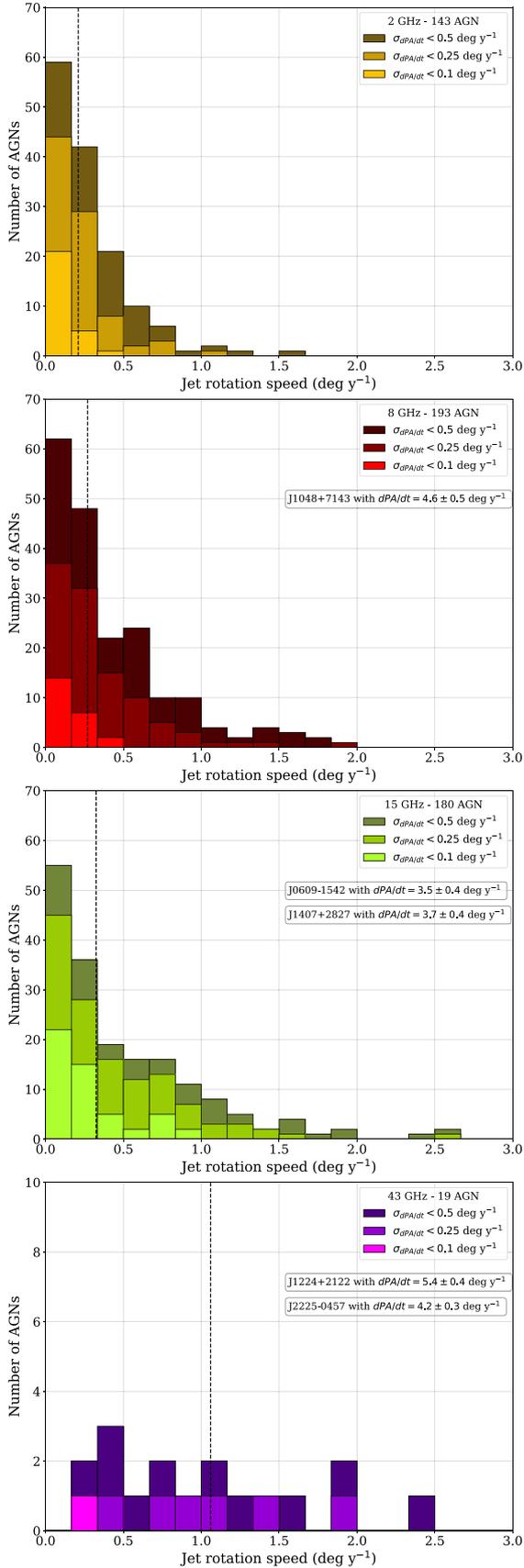

**Figure 5.** Distribution of AGNs by jet rotation speed as a function of frequency. Orange colour scheme represents the results at 2 GHz, red – 8 GHz, green – 15 GHz, violet – 43 GHz. The vertical dashed line indicates the median value of the rotation speed for each frequency.

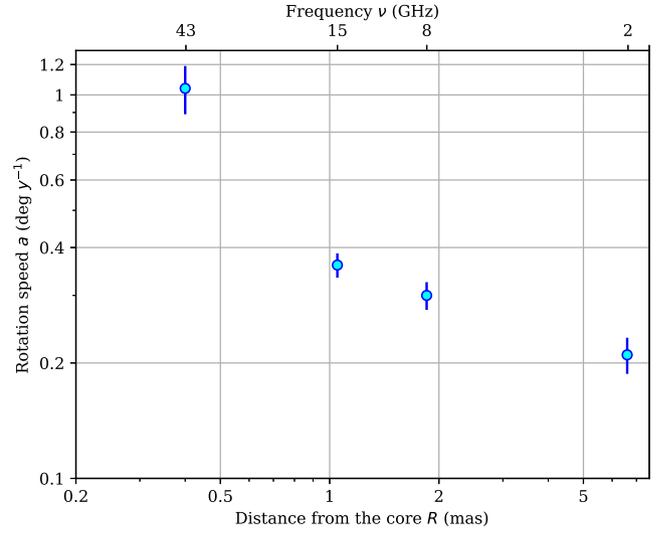

**Figure 6.** The median observed jet rotation speed as a function of the distance from the core. See Fig. 5 for the full distribution. Here, the decreasing trend along the jet is readily apparent (Section 4.1).

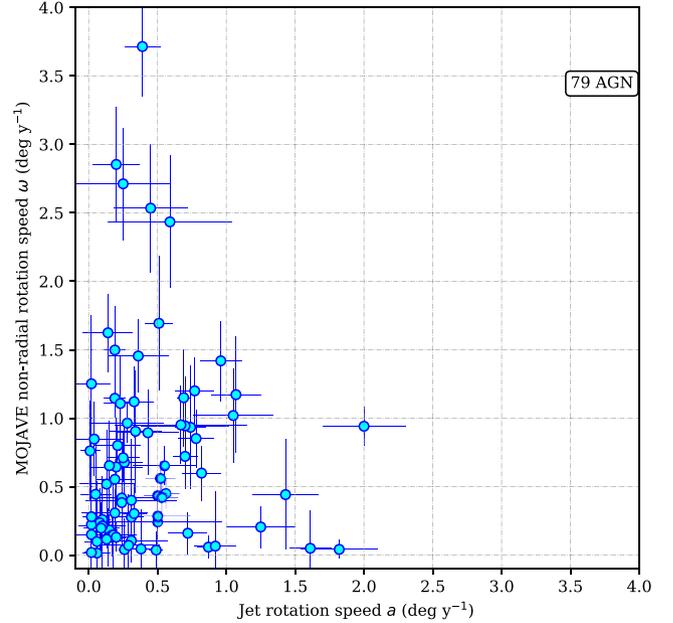

**Figure 7.** Relationship between the measured jet rotation speeds from the linear fit and the angular velocities of the non-radial motion of the individual components from the 15 GHz MOJAVE data (Section 4.2). This plot includes only those sources that are in both samples and have at least one MOJAVE component aside from the core. Note that both plot axes have the same scale; components can exhibit faster non-radial motion compared to apparent jet rotation speed, leading to no points in the right half of the plot.

### 4.4 Variability time-scales

We estimate the characteristic time-scale of jet direction changes from the following considerations. First of all, based on the available observational data, only a few sources exhibit complex rapid or discontinuous changes in PA on the observing time-scales. Thus, for most AGNs in the sample, their characteristic direction variation scales are longer than $T_{obs} \gtrsim 50$ years.

To constrain the time-scales from the other end, we note that the apparent rotation in the image cannot realistically cover more than half a circle in total. Among significantly variable jets in our analysis (Section 3), 90 per cent demonstrate rotation rates $a \geq 0.1$ deg yr$^{-1}$. These rates let us directly put an upper bound on the characteristic variability time-scales or potential periods: $T_{obs} \lesssim \pi/a \approx 2000$ yr.

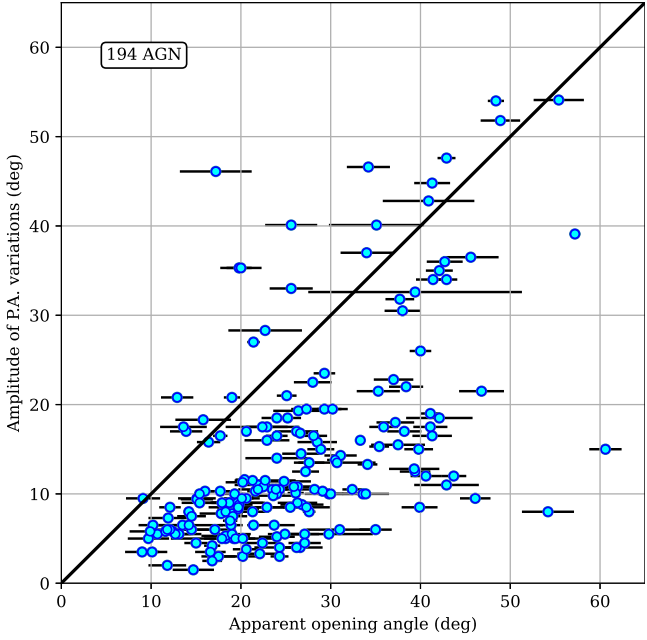

**Figure 8.** Relationship between the apparent opening angles of the jet at 15 GHz (Pushkarev et al. 2017) and the amplitude of jet PA variations. See Section 4.3 for motivation and discussion.

A more detailed consideration can be made on the basis of one of the existing models of jet geometry. Namely, the helical jet model, described in detail in Butuzova (2018) and Butuzova & Pushkarev (2020). The results of those studies also include a formula to describe the dependence of the PA of the jet at such a geometry, it has the form:

$$\text{PA}[\varphi(t)] = \pm \arctan \frac{\sin \varphi(t)}{\theta_0/\xi + \cos \varphi(t)} + \text{PA}_0.$$

The sign defines the direction of the jet rotation, $\varphi$ is the internal angle on the spiral, and the angles $\xi$ and $\theta$ are the jet half-opening and viewing angles respectively. PA$_0$ here, as above, is the average value of the jet PA around which the change of direction occurs.

For a uniform helix, the internal angle $\varphi$ changes linearly over time:

$$\varphi = \frac{2\pi t}{T_{obs}} + \varphi_0,$$

We focus on studying variability time-scales longer than the observation time. Then, the PA$(t)$ dependence can be well-approximated by a straight line, corresponding to a constant rotation rate. The most likely rotation rate, that is the most common $\frac{d\text{PA}(t)}{dt}$ over the whole period turns out to correspond to $\varphi_0 = 0, t = 0$. Its value is

$$\frac{d\text{PA}(t)}{dt} = \frac{2\pi}{T_{obs}(1 + \theta_0/\xi)}. \tag{1}$$

Hence, we obtain an expression for the period of variability, in the approximation of a constant rotation speed of the jet. According to Pushkarev et al. (2017), median value of the ratio $\theta_0/\xi$ for AGNs at 15 GHz is equal to $\theta_0/\xi \approx 2.5$. Thus, by expressing the period from this formula, we can obtain an upper bound on the potential variability period. Measured rotation rates $a \geq 0.1$ deg yr$^{-1}$ correspond to the potential periods being $T_{obs} \lesssim 1000$ yr.

As we see, both estimates on the period upper bound in this section produce comparable results. Further, we are going to use the slightly more conservative $T_{obs} \lesssim 2000$ yr bound. Note that these periods are in the reference frame of the observer. Meanwhile, the periods obtained from the consideration of physical precession scenarios correspond to the source reference frame. Therefore, we correct them by the redshift $z$:

$$T_{obs} = (1 + z) \cdot T_{intrinsic},$$

and obtain $T_{intrinsic} \lesssim 1000$ yr for a typical blazar at $z \sim 1$.

These characteristic time-scales are applicable only for the AGNs with direction variability detected in our study, that is a third of the sample. Some individual AGNs demonstrate more complex behaviour than our constant-rotation fits can capture (Fig. 4), but most of the remaining two thirds do not show significant direction variability within our uncertainties.

### 4.5 Scenarios causing variability

This section is devoted to discussing issues of possible scenarios of variability, including periodic and quasi-periodic variability. In this paper, we focus on the study and description of potentially periodic processes with periods equal to or longer than the entire observation time of individual active nuclei. Thus, the undoubtedly variable and potentially periodic PA behaviour of individual sources on scales of less than 10 yr, demonstrated by some quasars as, for example, in Fig. 4(f) remains beyond the scope of our study.

#### 4.5.1 Instabilities in the jet

Plasma instabilities in the bulk of the jet, or on the boundary with external environment, can cause jet wobbling and apparent direction variations. The interplay of different instabilities and corresponding stabilizing effects can produce a diverse set of effects in AGN jets (e.g. Perucho 2012). If these effects are indeed the cause of apparent direction variations, the decrease of variation amplitudes and speeds along the jet (Fig. 6) is consistent with instabilities on the jet boundary.

#### 4.5.2 Lense–Thirring effect

One of the possible scenarios for the occurrence of precession is the Lense–Thirring mechanism (e.g. Liu & Melia 2002; Caproni, Mosquera Cuesta & Abraham 2004). In this case, there is a disc-driven precession in which the change of direction of the relativistic ejection is connected with the precession of the central black hole because of different direction of its spin and angular momentum of the accretion disc surrounding it. For example, the precession periods estimated in Caproni et al. (2004) for the surface density that is $\propto r^{-1}$ are consistent with our estimated periods. Namely, they fall within the range from tens to thousands years, see Fig. 1 in Caproni et al. (2004). This agreement of time-scales makes the disc-related precession one of the feasible explanations for the observed jet direction variations.

#### 4.5.3 Binary systems

Under the assumption that the change in the direction of the jet is caused by the presence of a second black hole in the AGN, we can constrain the characteristic mass ratios of these two objects and the distance between them. The characteristic size of the binary system can vary over a very wide range. For example, according to An, Mohan & Frey (2018) or Severgnini et al. (2022), the distance between black holes can vary from 0.01 pc up to tens of pc.

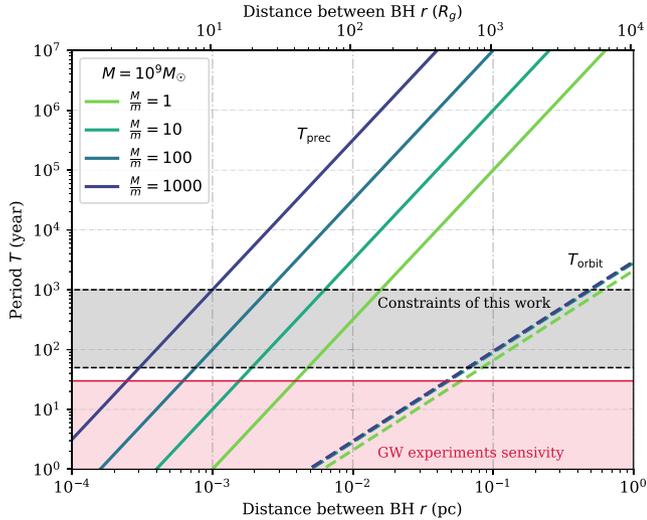

**Figure 9.** Connection between the period of the jet direction changes and the distance between the black holes. Shown are the scenarios of precession of the binary system (solid line) and of the orbital motion (dashed line). Different colours correspond to different mass ratios of the central and secondary black holes. The top axis corresponds to the Schwarzschild radius of the primary black hole. The mass of the primary black hole is assumed to be $M = 10^9 M_\odot$.

Let the mass of the primary black hole be $M$, the mass of the secondary black hole $m$, and the distance between them $r$. The causes of the change of the jet direction can be both the orbital motion of the secondary black hole around the central black hole and the induced precession of the axis of the central source, discussed in Begelman, Blandford & Rees (1980); Valtonen & Wiik (2012), and Dey et al. (2021).

To estimate the period of orbital motion, the classical mechanics is generally suitable beyond a few gravitational radii. For example, $T_{\rm orbit} \approx 300$ yr corresponds to $r \approx 0.2$ pc for $M \sim 10^9 M_\odot$. This distance-period relationship is shown in dashed lines in Fig. 9. Note that the distances larger than a tenth of a parsec or $\sim 10^3 R_g$ correspond to an angular size of $0.01\ldots0.1$ mas (at typical $z \sim 1$) and are within reach of direct detection by existing telescopes. For the moment, we consider such a scenario causing the observed variability in the apparent jet direction to be plausible. Still, future observations, including higher-resolution VLBI, are needed to unambiguously determine the presence or absence of such systems. Another clear indication of a binary system would be gravitational waves detections, but their characteristic frequencies would be measured in nanohertz, making them difficult to detect with current instruments.

Now consider the scenario of precession-driven variability described in Begelman et al. (1980). For the same values of the period and mass of the central black hole, the characteristic dependence of the precession period on the ratio of black hole masses is shown in Fig. 9. In this case, the characteristic distance between black holes in the binary system is thousandths of a parsec, or about $10 R_g$ of the central black hole.

Summarizing these two scenarios, we suggest that the orbital mechanism of the origin of the variability in the direction of the jet in the binary system is significantly more probable than the precession one. This is due to the fact that for the variability on time-scales estimated in Section 4.4 the characteristic distance between black holes should be extremely small, about $10 R_g$, and a stable long-term rotation of black holes in such binary systems is impossible, and moreover will require the effects of general relativity to be taken into account.

## 5 CONCLUSIONS

We have developed an automatic algorithm for determining the apparent inner jet direction on VLBI images of active galaxies, and applied it to archival images from the Astrogeo data base: 21 thousand AGNs observed between 1994 and 2023. We measured and investigated time dependencies of the jet PA in 317 AGNs utilizing 18 545 individual images in the frequency range from 2 to 43 GHz. Significant ($> 3\sigma$) jet direction variations were detected for 27 per cent AGNs in this sample. We believe the directional variability is a ubiquitous effect, potentially happening in the majority of AGNs with bright radio jets.

We focused on jet direction variations on time-scales longer than typical observing date ranges of 15–30 yr. Observationally, such variations appear as rotations at a constant speed. Average apparent rotation speed for frequently observed AGNs range from $0.21 \deg {\rm yr}^{-1}$ at 2 GHz (7 mas from the core) to $1.04 \deg {\rm yr}^{-1}$ at 43 GHz (0.4 mas from the core). There is a clear trend of increasing jet rotation speed with increasing frequency, when observations probe regions closer to the jet origin.

The strong rotation speed evolution along the jet indicates a nontrivial jet morphology. Indeed, in a ballistic propagation scenario with a precessing nozzle, the rotation speed would remain constant along the jet. Still, the jet direction changes are caused by the nozzle behaviour close to its origin, not by transverse bulk acceleration downstream.

The nozzle wobbling with required properties and time-scales of hundreds to thousands years can realistically be caused by several possible scenarios. They range from plasma instabilities within the jet to the accretion disc influence or binary system orbital motion in a binary system. We constrain their parameters whenever feasible: in particular, binary precession is highly unlikely due to very short component separation required; binary orbital motion, Lense–Thirring precession with the accretion disc density $\sim r^{-1}$, and plasma instabilities within the jet remain possible origins of jet wobbling.

Note that jet direction variations at different time-scales are studied best via different approaches. In this paper, scales longer than the observing time range are handled by linear fits of jet directions over time. Faster, shorter-scale variations can be studied through fitting more complicated quasi-periodical functions to direction measurements. Extremely long scales, more than tens of thousands of years, can be probed via indirect methods, such as kpc-scale helical structure.

### ACKNOWLEDGEMENTS

This study has been supported by the Russian Science Foundation: project 20-72-10078, https://rscf.ru/project/20-72-10078/. This work was supported by the Black Hole Initiative, which is funded by grants from the John Templeton Foundation (Grants 60477, 61479, and 62286) and the Gordon and Betty Moore Foundation (Grant GBMF-8273). The opinions expressed in this work are those of the authors and do not necessarily reflect the views of these foundations.

## DATA AVAILABILITY

The analysis is based on the VLBI observations compiled and publicly available in the Astrogeo[3] database.

## SUPPORTING INFORMATION

Supplementary data are available at *MNRAS* online.

**suppl_data**

Please note: Oxford University Press is not responsible for the content or functionality of any supporting materials supplied by the authors. Any queries (other than missing material) should be directed to the corresponding author for the article.

## APPENDIX A: SPECIAL CASES OF SHARP JET DIRECTION CHANGES

In Section 4.1, we discuss the typical behaviour of apparent jet direction in AGNs. Variability curves, typically demonstrate smooth variations well-described under the constant rotation speed assumption. However, sudden jumps and sharp changes may occur as well; we believe there are about 10 per cent of AGNs jets exhibiting such behaviour. In this appendix, we illustrate selected cases complex jet PA examples

As a visual aid, we present VLBI images from epochs during times of very fast PA changes in Fig. A1. These images are intended to help understanding how measured PA variations manifest themselves in terms of underlying radio emission structure detected with VLBI. Further, in Fig. A2, we highlight several jets with complex direction variability patterns.

J1224 + 2122 (4C 21.35). The jet rotation is highly non-uniform over time, becoming faster in the most recent epochs. Linear-fitting captures the overall trend, but fails with details.

J0854 + 2006 (OJ 287). A well-studied AGN, a prominent candidate for a binary black hole in the centre. The jet direction behaviour is complex at all frequencies and scales, making it one of the least good fits in the constant rotation speed assumption. Moreover, the jet exhibits a strong bending effect, leading to different PA values at different frequencies.

J0319 + 4130 (3C 84). The jet demonstrates a curved extended structure with highly pronounced edge brightening. The rapid changes in the apparent jet direction are partially caused by the brightest feature being at one edge or the other.

J0457 − 2324 (PKS 0454 − 234). The apparent jet direction significantly varies between observing frequencies, and visually we see is generally pointed southwards ($\approx 180°$). Erratic changes in the apparent direction are due to relatively low-brightness emission outside the core region.

J0730 − 1141 (PKS 0727 − 11). The jet PA is slightly variable in time and noticeably differs between observing frequencies. This behaviour is caused by the strongly curved structure of the jet, apparent in the images. We note that the extended

---

[3] http://astrogeo.org/vlbi_images/

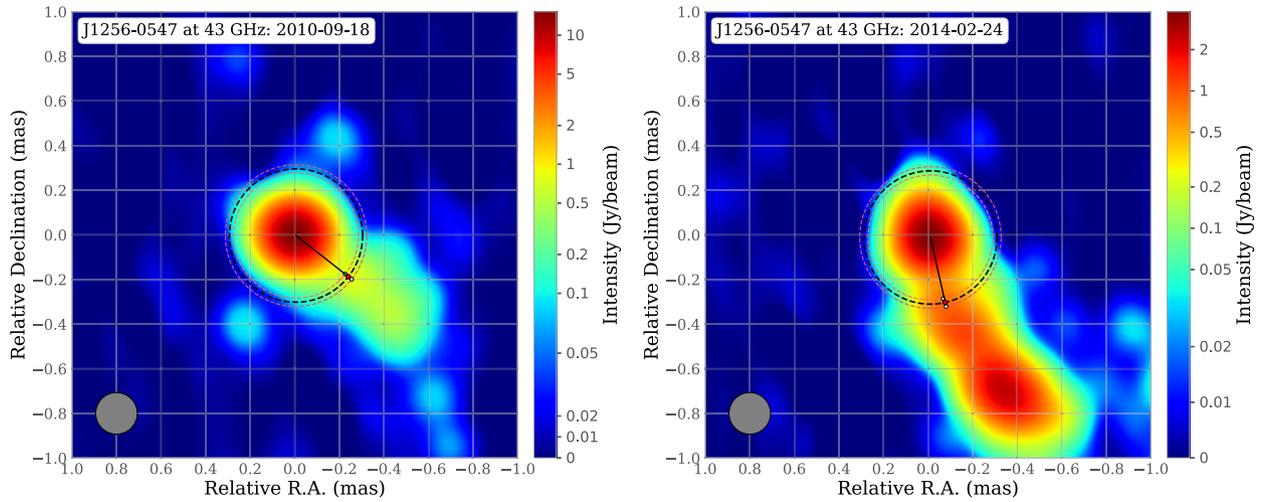

(a) VLBI images of 3C 279 at 43 GHz at September 18, 2010 and February 24, 2014. In those three and a half years, the jet abruptly changes its direction by more than 40°. See its variability curves in Figure 4.

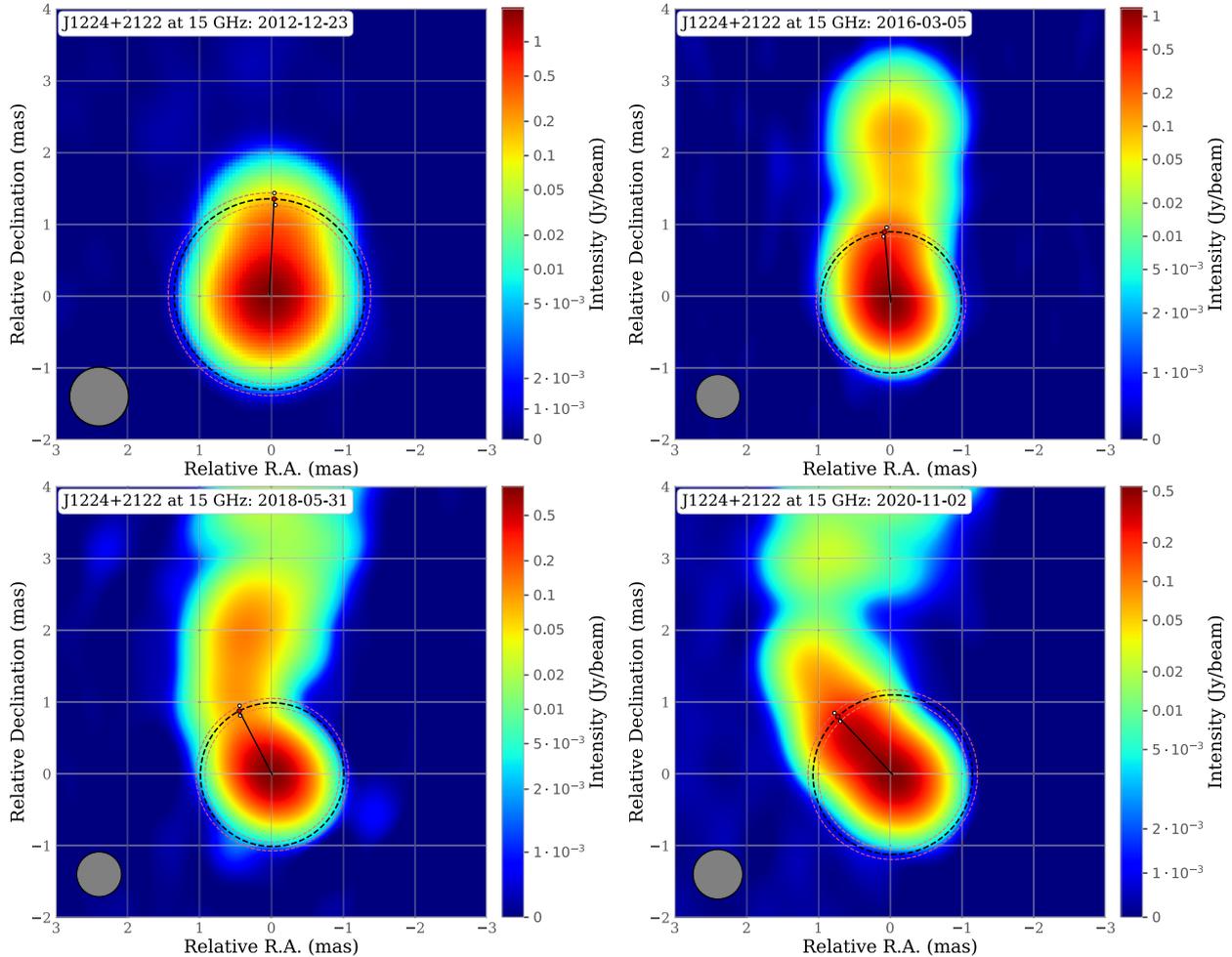

(b) J1224+2122 at 15 GHz show a significant and prolonged (about 8 years) counterclockwise rotation of the jet. See its variability curves in Figure A2.

**Figure A1.** Examples of VLBI images of AGN jets, together with their measured jet directions. The grey circles shows the size of the beam. These examples illustrate sudden changes in the apparent jet direction or fast jet rotations that are apparent in temporal plots (Fig. 4e).

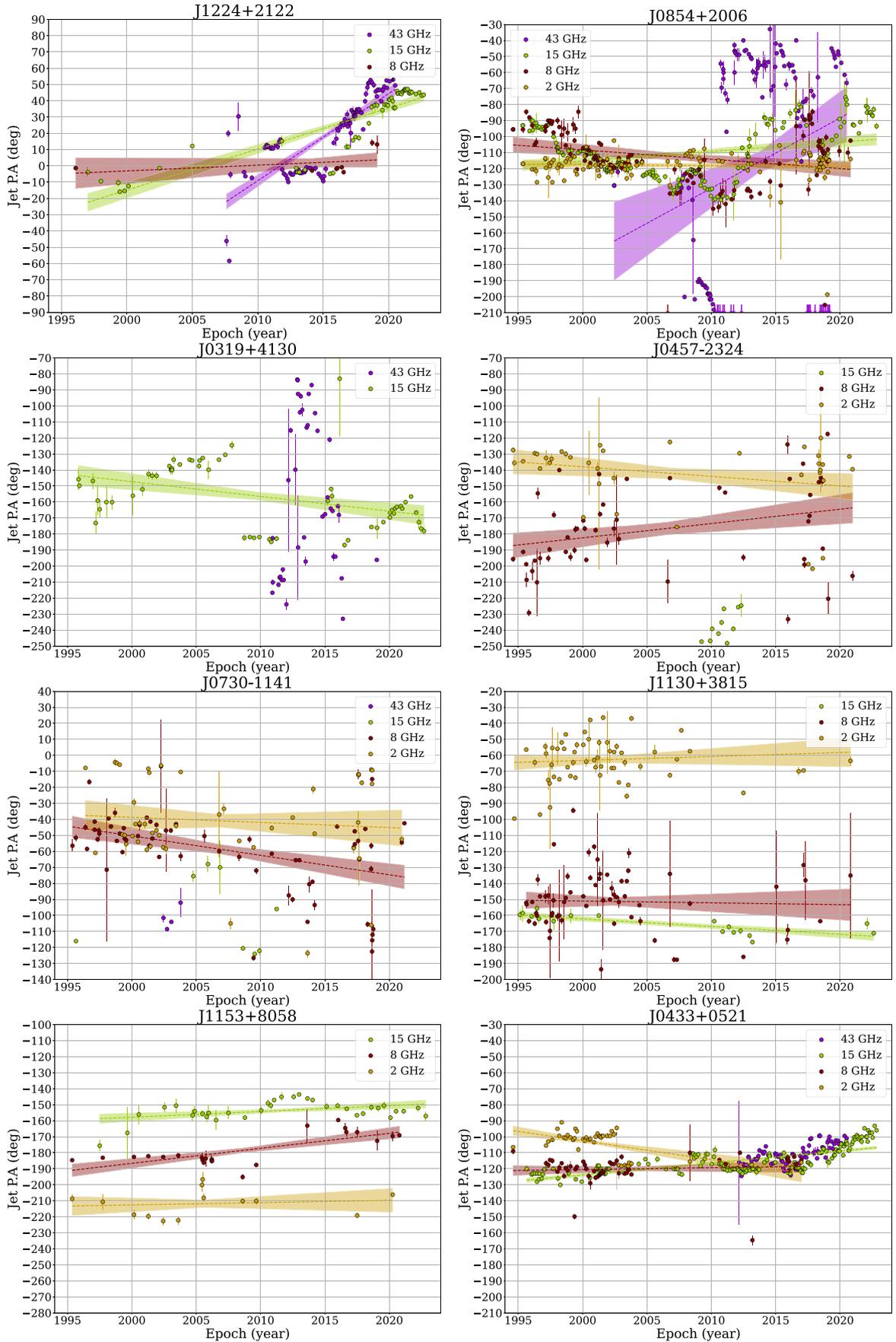

**Figure A2.** Jet PA variations over time for selected objects with large jet PA variations discussed in Appendix A. See Fig. 4 for plot elements explanation; note that the vertical axis limits are twice as large here compared to Fig. 4.

jet emission is also visible at 15 GHz, but measuring its PA at 1.05 mas (Table 1) is very challenging because of the patchy structure.

J1130 + 3815 (IVS B1128 + 385). The apparent jet direction changes significantly between observing frequencies. Visual inspection of the images indicates a highly curved structure at these scales.

J1153 + 8058 (S5 1150 + 81). Its jet demonstrates a curved structure, resulting in different PA at different frequencies. The slight change in direction at 15 GHz from 2010 to 2015 appears to be repeated at 8 GHz with some timelag.

J0433 + 0521 (3C 120). Its jet direction nearly follows linear trends at all frequencies, although with some inter-band variations. The jump at 2 GHz in 2003 is associated with the emergence of a new bright component around that time.

This paper has been typeset from a TEX/LATEX file prepared by the author.